\documentclass[preprint,proceedings]{rmaa}

\usepackage{amsmath}
\usepackage{verbatim}
\usepackage{mathrsfs}
\usepackage{amsfonts}
\usepackage{latexsym}
\usepackage{epsfig}
\usepackage{color}
\usepackage{graphicx,subfigure}
\usepackage{units}
\usepackage{dcolumn}
\usepackage{bm}
\usepackage{multirow}

\SetYear{2007}
\SetConfTitle{Name of Conference}

\title{On the possible enhancement of the Dark Matter density distribution at the Galactic Center }

\author{V. Gammaldi\altaffilmark{1,2,3,4}, V. Avila-Reese\altaffilmark{4}, O. Valenzuela\altaffilmark{4}, A. X. Gonzalez-Morales\altaffilmark{5}, P. Salucci\altaffilmark{1,2}, F. Nesti\altaffilmark{6}
}

\altaffiltext{1}{Scuola Internazionale Superiore di Studi Aavanzati, Italy;}\altaffiltext{2}{Istituto Nazionale di Fisica Nucleare, Trieste, Italy;}\altaffiltext{3}{Universidad Complutense de Madrid, Madrid, Spain;}\altaffiltext{4}{Universidad Nacional Aut\'onoma de M\'exico, M\'exico;}\altaffiltext{5}{CONACYT, Universidad de Guanajuato, Le\'on, M\'exico;}\altaffiltext{6}{Rudjer Boskovic Institute, Zagreb, Croatia.}

\suppressfulladdresses

\listofauthors{W. J. Henney}
\indexauthor{Henney, W. J.}

\addkeyword{H~II regions}
\addkeyword{ISM: Jets and outflows}
\addkeyword{Stars: Pre-main sequence}
\addkeyword{Stars: Mass loss}

\begin{document}
\maketitle 

\boldabstract{Una sobredensidad de materia oscura, de origen t\' ermico,  en el centro gal\' actico producida por la presencia del agujero negro Sgr A$^*$   podr\' ia explicar el factor astrof\' isico necesario para justificar el corte en el espectro de rayos gamma detectado por HESS.
\\ 
The Dark Matter (DM) spike induced by the adiabatic growth of a massive Black Hole (BH) in a cuspy environment, may explain the thermal DM density required to fit the cut-off in the HESSJ1745-290 $\gamma$-ray spectra (F. Aharonian et al. (2009)) as TeV DM signal with a background component (Cembranos et al. (2012)). The spike extension appears comparable with the HESS angular resolution.}The DM-density is locally enhanced in a region of radius $R_\text{sp}=\alpha_\gamma r_s(\text{M}_\text{BH}/\rho_s r_s)^{1/(3-\gamma)}$ as studied by Gondolo $\&$ Silk (1999) for several profiles ($\alpha_\gamma$). The BH mass at the GC is $\text{M}_\text{BH}=4.5\times 10^6\text{ M}_\odot$. We use the hydrodynamics Milky Way-like simulation Garrotxa (Roca-Fabregas et al. 2016). We fit the DM distribution to three cases (see for details Gammaldi et al. 2016): i) 4-parameter profile down to the nominal resolution limit of 109 pc (GARR-I), ii) 4-parameter profile with conservative limit of 300 pc (GARR-I300), and iii) 5-parameter profile from 300 pc (GARR-II300). The inner slopes, $\gamma$, which we extrapolate to the very center, are  0.6, 1 and 0.02, respectively. 
For the angular resolution of the HESS telescope ($\sim 0.1^\circ$) the astrophysical factor related with the BH induced DM spike on each profile is: $\langle J \rangle_{\Delta\Omega}^{\text{BH-GARRI}}=2.58\times10^{27}\text{GeV}^2\text{cm}^{-5}\text{sr}^{-1}$, $\langle J \rangle_{\Delta\Omega}^{\text{BH-GARRI300}}=2.16\times10^{27}\text{GeV}^2\text{cm}^{-5}\text{sr}^{-1}$ and $\langle J \rangle_{\Delta\Omega}^{\text{BH-GARRII}}=7.56\times10^{25}\text{GeV}^2\text{cm}^{-5}\text{sr}^{-1}$. Each one corresponds to $R_\text{sp}=16$ pc ($0.11^\circ$ deg),  $R_\text{sp}=11$ pc ($0.07^\circ$ deg) and $R_\text{sp}=2.3$ pc ($0.01^\circ$ deg) assuming $R_\odot=8.5$ kpc. The comparison of these results with the HESS data shows that the observed angular extent of the HESSJ1745-290 signal depends on not only the instrumental resolution, but also on the background normalization. In the upper panel of Fig.1 we assume that the background increases through the GC as the extrapolation of the underlying DM-halo profile without spike. 
\begin{figure}[!h]
\begin{center}\includegraphics[width=0.95\columnwidth]{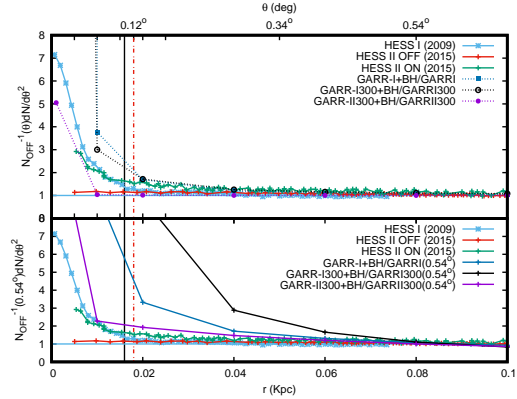}
\caption{\footnotesize{Angular distribution of the number of events of the HESSJ1745-290 data. Top: The expected DM-spike is normalized to a background model that is increasing through the GC. Down: The same when the background is assumed to be constant to its value at $\approx 80$ pc away fro the GC.}}
    \end{center}
\end{figure}
The case for GARR-I300 ($\gamma=1$) could be considered similar to the case in which the background is given by a millisecond pulsars (MSPs) population following the distribution of the GeV $\gamma$-ray emission claimed in Calore et al. (2015) (there, $\gamma=1.2$). 
In this case, the DM spike would appear much more localized than if the signal were normalized to the value of the background at $0.54^\circ$ deg ($\approx 80$ pc from the GC).
The DM spike may help to describe the spatial tail reported by HESS II at angular scales $0.54^\circ$ towards Sgr A$^*$. On the other hand, the different profiles of the spike may allow to make a difference to disentangle the nature (warm or cold) of the DM particle (Gammaldi, Nesti $\&$ Salucci, in preparation.).


%

\end{document}